\documentclass[11pt,a4paper]{article}
\usepackage{setspace}
\pdfoutput=1
\usepackage{latexsym,amssymb,amsmath,mathrsfs,amsfonts} 
\usepackage[retainorgcmds]{IEEEtrantools}
\usepackage{fullpage,natbib,color,
  graphics,graphicx,cases}

\makeatletter

\newcommand{\Rmnum}[1]{\expandafter\@slowromancap\romannumeral #1@}
\newcommand{\mbbR}{\mbox{$\mathbb{R}$}}

\newcommand{\mbbI}{\mbox{$\mathbb{I}$}}

\newcommand{\mbbN}{\mbox{$\mathbb{N}$}}

\newcommand{\Beta}{\mbox{Be}}

\makeatother

\title{Extended Generalised Pareto Models for Tail Estimation}

\author{Ioannis Papastathopoulos\footnote{Department of Mathematics
    and Statistics, Lancaster University, Lancaster, LA1 4YF,
    UK. Email: i.papastathopoulos@lancaster.ac.uk and
    j.tawn@lancaster.ac.uk} ~and Jonathan A. Tawn}
\date{\today}

\newtheorem{theorem}{Theorem}

\newtheorem{corollary}{Corollary}

\definecolor{light-grey}{gray}{0.80}
\begin{document}
\maketitle
\begin{abstract}
  The most popular approach in extreme value statistics is the
  modelling of threshold exceedances using the asymptotically
  motivated generalised Pareto distribution. This approach involves
  the selection of a high threshold above which the model fits the
  data well. Sometimes, few observations of a measurement process
  might be recorded in applications and so selecting a high quantile
  of the sample as the threshold leads to almost no exceedances. In
  this paper we propose extensions of the generalised Pareto
  distribution that incorporate an additional shape parameter while
  keeping the tail behaviour unaffected. The inclusion of this
  parameter offers additional structure for the main body of the
  distribution, improves the stability of the modified scale, tail
  index and return level estimates to threshold choice and allows a
  lower threshold to be selected. We illustrate the benefits of the
  proposed models with a simulation study and two case studies.
\end{abstract}

\section{Introduction}
\label{sec:intro}
The area of extreme value theory focuses on the study and development
of stochastic models that can be used for inference on applied
problems related to the frequency of very big (or very small) values
in random experiments. One such widely used model is the generalised
Pareto (GP) distribution defined by its distribution function
\begin{equation}
  F(x;\boldsymbol{\lambda}) = 1-\left(1 + \xi x/\sigma\right)_{+}^{-1/\xi},
  \quad x>0,
  \label{eq:gpd}
\end{equation}
where $\boldsymbol{\lambda}=(\sigma, \xi)$ is a vector of parameters
in $(0,\infty)\times (-\infty,\infty)$ and $z_{+}=\max(z,0)$. Consider
a random variable $X$ arising from an absolutely continuous
distribution function $F_X$ and let also $x^{F_X} =
\sup\{x:F_X(x)<1\}$ be the upper end point of $F_X$.\ \cite{pick75}
shows that if there exists a scaling function $h_X(u):\mbbR
\rightarrow \mbbR_+$, $u<x^{F_X}$, such that the scaled excess random
variable $\{(X-u)/h_X(u)\}|X>u$ converges in distribution to a
non-degenerate limit as $u\rightarrow x^{F_X}$, then this is
necessarily of the same type as the GP distribution, i.e.,
\begin{equation}
  \lim_{u\rightarrow x^{F_X}} \Pr\left\{\frac{X-u}{h_X(u)} < x \Big |
    X>u\right\} = F(x;\boldsymbol{\lambda}_0), \quad x>0,\quad \boldsymbol{\lambda}_0=(1,\xi).
  \label{eq:gpdlimit}
\end{equation}
\noindent
Without loss of generality, the scaling function $h_X$ can be defined
by the reciprocal hazard function of $X$, i.e.,
$h_X(u)=\{1-F_X(u)\}/F_X'(u)$. \cite{pick86} shows that a necessary
and sufficient condition for twice differentiable convergence is
$\lim_{u\rightarrow x^{F_X}}h_X'(u)=\xi$, meaning that not only
limit~(\ref{eq:gpdlimit}) holds but the corresponding densities and
derivatives of densities converge. The parameter $\xi$ is most
commonly referred to as the shape parameter or the tail index of the
distribution and we adopt the latter since we use the word shape for a
different characteristic in the paper. The sign of the tail index
$\xi$ indicates the decay of the tail of $F_X$: $\xi>0$ means that
$F_X$ has a heavy-tailed distribution, $\xi\rightarrow 0$ corresponds
to exponential decay and $\xi<0$ means that $F_X$ has finite upper end
point, i.e., $x^{F_X}<\infty$. \newline

\noindent Consider a sequence of independent and identically
distributed measurements $\{x_1,\hdots,x_n\}$ arising from the random
variable $X$. Standard practice in applications where we want to
estimate extreme quantiles of the underlying distribution is to follow
the approach of \cite{davismit90} and assume that the limit
relationship~(\ref{eq:gpdlimit}) holds exactly for some threshold
$u<\max_{i=1,\hdots,n}(x_i)$; i.e., $X-u|X>u$ has distribution
function $F(x;\sigma_{u},\xi)$ where the function $h_X(u)$ is absorbed
in the distribution function $F$ as a threshold dependent scale
parameter $\sigma_u>0$. Extreme events are defined by the threshold
exceedances $\{x_i: x_i>u\}$ and subsequently, the GP distribution is
fitted to the random sample of excesses $\{x_i-u: x_i>u\}$ using
maximum likelihood techniques. The fitted model is then extrapolated
to levels above which no data are observed. Two central assumptions
are imposed when using this procedure in practice. One is that the
asymptotic argument of equation~(\ref{eq:gpdlimit}) is valid for the
distribution function of the data under study and second is that an
appropriate threshold $u$ can be found such that the GP model provides
a good approximation to exceedances of $u$. For a number of reasons
such as the cost and time of collecting data, few observations of a
measurement process might be recorded in applications. Since
$u<\max_{i=1,\hdots n}(x_i)$ and $n$ is small, $F_X(u)\ll 1$ and so
limit~(\ref{eq:gpdlimit}) is likely to be a poor approximation to the
distribution of exceedances of $u$ in such cases.\newline

\begin{figure}[!htbp]
  \centering
  \includegraphics[height=9.5cm,width=9.5cm,trim= 20 15 20 20]{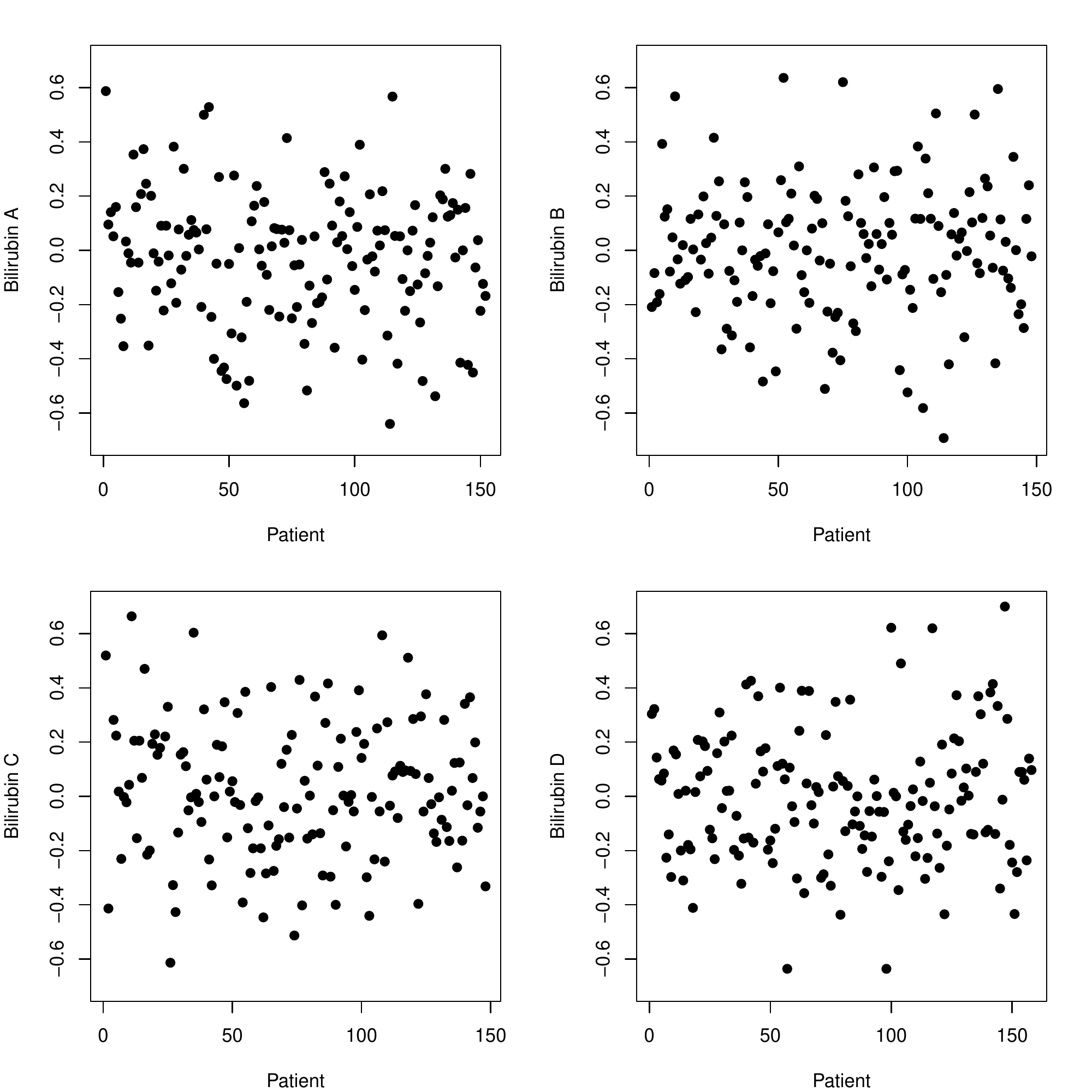}
  \caption{Residual total bilirubin variable measured from 606
    patients at four different doses $A$, $B$, $C$ and $D$.}
  \label{fig:pharma_data}
\end{figure}

\noindent Figure~\ref{fig:pharma_data} shows such an example of
residual bilirubin data collected from a clinical study of 606
patients who were randomized to 4 doses of a drug, the highest dose of
which is considered to have potential for liver toxicity. Data were
available prior to treatment (baseline visit) and after 6 weeks of
treatment (postbaseline). The residual bilirubin observations are the
residuals of linear median regression models of postbaseline on the
baseline in each dose, see also \cite{texmex} for a similar
analysis.\newline

\noindent
The 4 different doses are coded here with increasing dose as $A$, $B$,
$C$ and $D$. The published literature reports jaundice, hepatitis and
similar symptoms in approximately 1 out of 500 patients taking the
dose $D$ of this drug, see \cite{texmex}. According to \cite{FDAliver}
joint occurrence of extremes of the total bilirubin and
aminotransferase laboratory variables are indicative of drug induced
liver toxicity. Therefore, proper statistical modelling of their
extremes is vital for assessing the liver toxicity of a new
drug. However, the limited amount of information in each dose
illustrates the problems of relying on the GP distribution. We will
return to the analysis of these data in \S~\ref{sec:pharma}. \newline

\noindent The issue of specifying an appropriate threshold for fitting
the GP distribution constitutes the major problem of the
\cite{davismit90} approach.\ Typically, a threshold $u$ is chosen as
the lowest possible value above which the estimates of the tail index
$\xi$ and the modified scale $\sigma^{*}=\sigma_u-\xi u$ stabilise
\citep[see][]{cole01}.\ Departures from the GP distribution again
imply that such a threshold might not be observable. Moreover, the
higher the threshold the larger the sampling variability of the
estimates of the modified scale and tail index parameters which leads
to estimates being unstable over different thresholds. A variety of
methods have been developed in the literature to address the problem
of departures from the GP model assumption. \cite{peng98},
\cite{feuerhall99} and \cite{beiretal99} among others, use second
order refined models to select the threshold and proceed with the
estimation of tail characteristics by using the GP distribution.
\cite{beiretal09} also use a second order approach for the modelling
of the tail probability as well as for the extrapolation. An even more
unsettling feature arises in cases where two or more seemingly
plausible thresholds yield significantly different estimates of
extreme quantities of interest. In such circumstances, the threshold
selection is liable to be the subjective choice of the
practitioner. \cite{frigessi02} propose unsupervised tail estimation
with the use of a dynamic mixture model aimed for the entire
distribution of the data. \cite{tancredi06} take into account the
threshold uncertainty by allowing the threshold to be estimated as a
statistical parameter in a mixture model.\ \cite{macscar11} also use a
mixture model where the non-extreme part of the data is estimated
non-parametrically. \cite{wadtawn11} exploit penultimate theory to
model threshold uncertainty and provide a likelihood ratio testing
procedure for the threshold selection. \newline

\noindent
All of the aforementioned approaches are either based on second order
asymptotic arguments adding a little extra flexibility to the fit of
the GP distribution or they model the entire distribution, i.e., model
the body as well as the tail. Furthermore, some of these approaches
are confined to the heavy-tailed case $\xi>0$ \citep{beiretal09}. Our
goal in this article is to construct parametric models for exceedances
over thresholds that are more flexible than the GP distribution and
add further insight on the threshold selection problem. For this
purpose, in \S~\ref{sec:hazard} we construct a new class of
probability models with distribution function
$G(x;\kappa,\boldsymbol{\lambda})$, where $\kappa>0$ is a shape
parameter, that generalises the GP distribution in the sense that
there exists a $\kappa^{*}>0$ such that
$G(x;\kappa^{*},\boldsymbol{\lambda})=F(x;\boldsymbol{\lambda})$. This
parameter $\kappa$ offers additional structure by inducing skewness to
the GP distribution while retaining $\xi \in \mbbR$ as the tail index.
Thus, a class of departures from the GP assumption of
limit~(\ref{eq:gpdlimit}) are captured by this shape parameter
$\kappa$. We will show that the inclusion of $\kappa$ improves the
stability of the estimates of the tail index and modified scale
parameter and allows a lower threshold to be selected. Consequently,
extrapolations based on different thresholds are more stable than
those obtained from the GP distribution, a feature which makes the
choice of threshold less important.  \newline

\noindent
In \S~\ref{sec:theory} we present three extensions of the GP
distribution and derive a characterisation of a class of models which
includes these examples.\ The new models are given along with the
description of the methodology implemented for the analysis of
threshold exceedances. A statistical test aiding threshold selection
is also illustrated. The effect of the new probability models on the
statistical analysis of extremes is assessed with a simulation study
in \S~\ref{sec:simulation}. Finally, in \S~\ref{sec:apps} we
illustrate the benefits of the new models through the analysis of
extreme flow data of the River Nidd, a dataset with known difficulties
in threshold selection, and the clinical trial data in
Figure~\ref{fig:pharma_data}.

\section{Theory and Models}
\label{sec:theory}

\subsection{Notation and motivation}
\label{sec:notmot}

Here and throughout, we denote
by $\beta(\cdot;a,b)$ and $\gamma(\cdot;a)$ the regularised incomplete
beta and regularised lower incomplete gamma functions given by
\begin{IEEEeqnarray*}{rCl}
  \beta(x;a,b) &=& \frac{1}{\Beta(a,b)} \int_{0}^{x} t^{a-1}
  (1-t)^{b-1} dt, \quad0 \leq x \leq 1, \\\\
  \gamma(y;a) &=& \frac{1}{\Gamma(a)} \int_{0}^{y} t^{a-1} e^{-t} dt,
  \quad 0 \leq y < \infty,
\end{IEEEeqnarray*}
with $a,b>0$. We also denote by $\beta^{-1}(\cdot,a,b)$ and
$\gamma^{-1}(\cdot,a)$ their corresponding inverses.
\newline

\noindent
Our examples are motivated by transformations of the form
$W:=F^{-1}(V;\boldsymbol{\lambda})$, where $F$ is the GP distribution
function given by equation~(\ref{eq:gpd}) and $V$ is a random variable
with support the unit interval $\mbbI=[0,1]$. The distribution
function and density function of $W$ are denoted by $G$ and $g$,
respectively. If $V\sim$~uniform(0,1) distribution then
$W\sim$~GP$(\sigma,\xi)$ but if $V$ is on $\mbbI$ with a distribution
that contains the uniform$(0,1)$ as a special case then more flexible
distributions than the GP are produced for $W$.

\subsection{Probability integral transform and new models}
\label{sec:pit}
\noindent
Let $\Omega_{Y}$, $\Omega_{V}$ be two sample spaces. Define
$Y:\Omega_{Y} \rightarrow \mbbR$ to be a random variable with
continuous probability density (distribution) function
$f_{Y}(x;\boldsymbol{\eta})$ $\left(F_{Y}(x;\boldsymbol{\eta})\right)$
parametrised over an $m$-dimensional vector of parameters
$\boldsymbol{\eta} \in \mbox{H}\subseteq\mbbR^m$. Let also $V:\Omega_{V}
\rightarrow \mbbR$ be a random variable with continuous probability
density (distribution) function $f_{V}(v;\boldsymbol{\theta})$
$(F_{V}(v;\boldsymbol{\theta}))$ parametrised over a $d$-dimensional
vector of parameters $\boldsymbol{\theta}\in \Theta \subseteq
\mbbR^d$, with $f_{V}(v;\boldsymbol{\theta})=0$, if $v\notin
[0,1]$. We also assume the existence of a $\boldsymbol{\theta}^{*} \in
\Theta$ such that $f_{V}(v;\boldsymbol{\theta}^{*})=1$, $\forall v\in
[0,1]$, i.e., a special case of $V$ follows the uniform(0,1)
distribution.\ Then, the distribution and density functions of the
transformed random variable $F_{Y}^{-1}(V;\boldsymbol{\eta})$ are
given by
\begin{IEEEeqnarray}{rCl}
  K(x;\boldsymbol{\eta},\boldsymbol{\theta}) &=&
  \Pr\left\{F_Y^{-1}(V;\boldsymbol{\eta})\leq x\right\} =
  F_{V}\left\{F_Y(x;\boldsymbol{\eta});\boldsymbol{\theta}\right\},\label{eq:gen}\\
  \nonumber \\ k(x;\boldsymbol{\eta},\boldsymbol{\theta})&=&
  K'(x;\boldsymbol{\theta},\boldsymbol{\eta})=f_{V}
  \left\{F_Y(x;\boldsymbol{\eta});\boldsymbol{\theta}\right\}f_Y(x;\boldsymbol{\eta}).
  \label{eq:gen_dens}
\end{IEEEeqnarray}  
Therefore the distribution function
$K(x;\boldsymbol{\eta},\boldsymbol{\theta})$ is a generalised
distribution function for $F_Y(x;\boldsymbol{\eta})$ in the sense that
$K(x;\boldsymbol{\boldsymbol{\eta},\theta}^{*})=F_Y(x;\boldsymbol{\eta})$,
i.e.,\ the distribution function $F_Y(x;\boldsymbol{\eta})$ is a
special case of $K(x;\boldsymbol{\eta},\boldsymbol{\theta})$.\newline

\noindent
Equations~(\ref{eq:gen}) and~\ref{eq:gen_dens} provide the basis of
all subsequent generalizations we propose. The distribution function
of $V$ can be constructed in several different ways, one of which is
the composition of a distribution function
$L_1(\cdot;\boldsymbol{\psi})$, $\boldsymbol{\psi}\in \Psi \subseteq
\mbbR^{d-m}$, $\dim(\boldsymbol{\psi})=d-m$, $d>2m$, with the inverse
of a distribution function $L_0(\cdot;\boldsymbol{\eta})$,
$\boldsymbol{\eta}\in \mbox{H} \subseteq \mbbR^{m}$,
$\dim(\boldsymbol{\eta})=m$, where $L_0$ and $L_1$ are defined on the
same support and $L_0$ is a special case of $L_1$ ($d>2m$ is required
for $\boldsymbol{\eta}$ to have lower dimension than
$\boldsymbol{\psi}$), that is
\begin{equation}
  F_{V}(v;\boldsymbol{\theta})=
  \begin{cases}
    0 & v < 0,\\
    L_1  \left\{L_0^{-1} \left(v;\boldsymbol{\eta}\right);\boldsymbol{\psi}\right\} & 0 \leq v \leq 1,\\
    1 & v>1,
  \end{cases}
  \label{eq:composition}
\end{equation}
where $\boldsymbol{\theta}\in \Theta \subseteq \mbbR^d$ consists of
elements taken from the combined vector
$(\boldsymbol{\psi},\boldsymbol{\eta})$. When $\boldsymbol{\psi}$ and
$\boldsymbol{\eta}$ have elements in common, the combined vector is
interpreted as the vector consisting of the unique elements of
$(\boldsymbol{\psi},\boldsymbol{\eta})$ that span $\Theta$. Note here
that $\boldsymbol{\theta}$ is not necessarily equal to the combined
vector $(\boldsymbol{\psi},\boldsymbol{\eta})$ since common elements
of $\boldsymbol{\psi}$ and $\boldsymbol{\eta}$, if any, are allowed to
cancel in composition~(\ref{eq:composition}). For instance, when $L_1$
and $L_0$ are the distribution functions of the gamma$(\kappa,\sigma)$
and exponential$(\sigma)$ random variables, $\kappa>0$, $\sigma>0$,
i.e., $L_1(x;\boldsymbol{\psi})=\gamma(x/\sigma,\kappa)$ and
$L_0(x;\boldsymbol{\eta})=1-\exp\left\{-x/\sigma\right\}$, for $x>0$,
then equation~(\ref{eq:composition}) yields the distribution function
$F_{V}(v;\boldsymbol{\theta})=\gamma\{-\log(1-v),\kappa\}$, $0\leq v
\leq 1$. Here
$\{\boldsymbol{\psi},\boldsymbol{\eta},\boldsymbol{\theta}\}=\{(\kappa,\sigma),\sigma,
\kappa\}$ and $\{d,m,\dim (\Theta)\}=\{3,1,1\}$. Moreover, this
distribution function reduces to the uniform$(0,1)$ distribution when
$\kappa=1$, i.e., $\boldsymbol{\theta}^{*}=1$.  \newline

\noindent
Below we present three new probability density functions that are
generalisations of the GP density and can be obtained by
transformations of the form $F_Y^{-1}(V;\boldsymbol{\eta})$, where
$F_Y=F$, $\boldsymbol{\eta}=\boldsymbol{\lambda}$ and $V$ is a random
variable that satisfies equation (\ref{eq:composition}). Owing to the
fact that each model extends the GP distribution in a parametric
fashion, we refer to the new models as the extended GP (EGP) models
and denote their density function by
$g(x;\boldsymbol{\lambda},\boldsymbol{\theta})$, for $x>0$.
\begin{description}
\item[Example 1] Let
  $F_{V}(v;\boldsymbol{\theta})=\beta\left\{1-(1-v)^{|\xi|},\kappa,|\xi|^{-1}\right\}$,
  $\boldsymbol{\theta}=(\kappa,\xi) \in
  (0,\infty)\times(-\infty,\infty)$. Then the transformed random
  variable $F^{-1}(V;\boldsymbol{\lambda})$ has probability density
  function given by
  \begin{numcases}{g(x;\boldsymbol{\lambda},\boldsymbol{\theta})=}
    \frac{|\xi|/\sigma}{\Beta(\kappa,|\xi|^{-1})}
    \left\{1-\left(1+\xi
        x/\sigma\right)_{+}^{-|\xi|/\xi}\right\}^{\kappa-1}
    \left(1+\xi x/\sigma\right)_{+}^{-1/\xi-1} & $\xi \neq 0$,
    \nonumber \\      \label{eq:EGP1} \\
    \frac{\sigma^{-1}}{\Gamma(\kappa)}x^{\kappa-1}e^{-x/\sigma}&
    $\xi \rightarrow 0$. \nonumber
  \end{numcases}

\item[Example 2] Let
  $F_{V}(v;\boldsymbol{\theta})=1-\gamma\left\{-\log
    (1-v),\kappa\right\}$,
  $\boldsymbol{\theta}=\kappa\in(0,\infty)$. Then the transformed
  random variable $F^{-1}(V;\boldsymbol{\lambda})$ has probability
  density function given by
  
  \begin{numcases}{g(x;\boldsymbol{\lambda},\boldsymbol{\theta})=}
    \frac{\sigma^{-1}}{\Gamma(\kappa)}\left\{\xi^{-1}\log
      \left(1+\xi x/\sigma\right)_{+} \right\}^{\kappa-1}
    \left(1+\xi x /\sigma\right)_{+}^{-1/\xi-1} & $\xi \neq 0$,
    \nonumber \\      \label{eq:EGP2} \\
    \frac{\sigma^{-1}}{\Gamma(\kappa)}x^{\kappa-1}e^{-x/\sigma}&
    $\xi \rightarrow$ 0.\nonumber
  \end{numcases}
  
\item[Example 3] Let $F_{V}(v)=v^{\kappa}$,
  $\boldsymbol{\theta}=\kappa\in(0,\infty)$. Then the transformed
  random variable $F^{-1}(V;\boldsymbol{\lambda})$ has probability
  density function given by
  \begin{numcases}{g(x;\boldsymbol{\lambda},\boldsymbol{\theta})=}
    \frac{\kappa}{\sigma}\left\{1-\left(1+\xi
        x/\sigma\right)_{+}^{-1/\xi}\right\}^{\kappa-1} \left(1+\xi x
      /\sigma\right)_{+}^{-1/\xi-1} & $\xi \neq 0$,
    \nonumber \\   \label{eq:EGP3}  \\
    \frac{\kappa}{\sigma}\left(1-e^{-x/\sigma}\right)^{\kappa-1}e^{-x/\sigma}&
    $\xi \rightarrow 0$.\nonumber
  \end{numcases}
\end{description}
We write $W\sim$~EGP1$(\kappa,\sigma,\xi)$,
$W\sim$~EGP2$(\kappa,\sigma,\xi)$ and
$W\sim$~EGP3$(\kappa,\sigma,\xi)$ when the density of a random
variable $W$ is given by expression~(\ref{eq:EGP1}), (\ref{eq:EGP2})
and (\ref{eq:EGP3}) respectively. In all examples in addition to the
GP parameters $\sigma$ and $\xi$ there is a shape parameter $\kappa>0$
that adds more flexibility in the main body of the density and does
not alter its tail behaviour, i.e., all distributions have tail index
$\xi\in\mbbR$. \newline

\noindent All models reduce to the GP density when $\kappa=1$. More
specifically, the EGP1 model can be viewed as an extended Snedecor's
$F_{\nu_1,\nu_2}$ distribution \citep{abrasteg65} with parameters
$\nu_1>0$ and $\nu_2 \in \mbbR$, that allows for negative $\nu_2$
giving finite upper bound for this distribution. When $\xi>0$ and
$\kappa=\sigma$, the density reduces to the $F_{\nu_1,\nu_2}$
distribution with $\nu_1=2\kappa$ and $\nu_2=2/\xi$. Additionally for
$\xi>0$, the EGP1 model is a well used loss distribution in actuarial
science known in that literature as the \textit{generalised Pareto}
distribution \citep{hoggklug84,klugetal04}. The EGP1 model is an
extension of this loss distribution for the case $\xi<0$. The EGP2
model can be viewed as a model that generalises the GP density in a
similar way to the gamma generalising the exponential
distribution. Specifically, the GP distribution is the distribution of
the random variable $(e^{\xi Y}-1)\sigma/\xi$, where $Y$ follows the
exponential(1) distribution \citep{hoskwallis87}. Analogously, the
EGP2 model is the distribution of the random variable $(e^{\xi
  Z}-1)\sigma/\xi$, where $Z$ follows the gamma($\kappa$,1)
distribution.\ Finally, the EGP3 distribution function is simply
obtained by raising the GP distribution function
$F(x;\boldsymbol{\lambda})$ to a power $\kappa>0$.

\subsection{Construction of extreme value models}
\label{sec:hazard}
Expression~(\ref{eq:composition}) represents a class of distribution
functions $F_{V}$. However, unlike the extended models of
\S~\ref{sec:pit}, the transformation
$F_Y^{-1}(V;\boldsymbol{\lambda})$ does not always ensure that the
resulting random variable has a tail index $\xi$ for all values of
$\boldsymbol{\theta}$. One such example can be obtained by taking
$F_V=L_1\circ L_0^{-1}$, with $L_1$ and $L_0$ being the distribution
functions of Weibull$(\kappa,\sigma)$ and exponential$(\sigma)$ random
variables, $\kappa$, $\sigma>0$, i.e.,
$L_1(x)=1-\exp\left\{-\left(x/\sigma\right)^\kappa\right\}$. In this
case, the transformed variable $F^{-1}(V;\boldsymbol{\lambda})$, has
survival function $\bar{G}$ given by
\[
\bar{G}(x;\kappa,\boldsymbol{\lambda}) = \exp\left[-
  \left\{\xi^{-1}\log\left(1+ \xi
      x/\sigma\right)_{+}\right\}^\kappa\right], \quad x>0
\]
which is a slowly varying function at $\infty$ for $\kappa<1$, $\xi>0$
and is therefore considered to be a `super-heavy-tailed' distribution
under this combination of parameters which means that the parameter
$\xi$ is no longer the tail index of this distribution. Hence, we
proceed b characterising in Theorem~\ref{th:th1} the class of
distribution functions $F_{V}$ under the assumption that
$F_Y^{-1}(V;\boldsymbol{\eta})$ has tail index $\xi \in \mbbR$ for all
values of $\boldsymbol{\theta}$.
\begin{theorem}
  \label{th:th1}
  Let $d>2m$ where $d,m \in\mbbN$ and consider the parameter vectors
  $(\boldsymbol{\eta},\boldsymbol{\psi},\boldsymbol{\theta})\in
  (\mbox{\textnormal{H}} \times \Psi \times \Theta) \subseteq
  (\mbbR^{m} \times \mbbR^{d-m} \times \mbbR^{d})$ with
  $\dim(\boldsymbol{\eta})=m$ and $\dim(\boldsymbol{\psi})=d-m$. Let
  $F_Y(x;\boldsymbol{\eta})$ be a twice differentiable distribution
  function of a random variable $Y$ admitting a density function
  $f_Y(x;\boldsymbol{\eta})$. Let also $V$ be a random variable with
  twice differentiable distribution function
  $F_{V}(v;\boldsymbol{\theta})$ so that its density function
  satisfies $f_{V}(v;\boldsymbol{\theta})=0$ when $v\notin [0,1]$.\
  Then the transformed random variable $F_Y^{-1}(V;\boldsymbol{\eta})$
  has tail index $\xi \in \mbbR$, if and only if, the distribution
  function of $V$ can be represented by
  \begin{equation}
    F_{V}(v;\boldsymbol{\theta}) = 1-\exp\left\{-\int_{0}^{v} \frac{dz}
      {\xi f_Y\{F_Y^{-1}(z);\boldsymbol{\eta}\} 
        C(z;\boldsymbol{\eta},\boldsymbol{\psi}) }\right\},\quad 0 \leq v \leq 1,
    \label{eq:Vdf}
  \end{equation}
  where $C(z;\boldsymbol{\eta},\boldsymbol{\psi})=\int
  [s(z;\boldsymbol{\eta},\boldsymbol{\psi})/f_Y\{F_Y^{-1}(z);\boldsymbol{\eta}\}]
  dz$ and $s$ is a real-valued function with \newline$\lim_{z
    \rightarrow 1} s(z;\boldsymbol{\eta},\boldsymbol{\psi}) = 1$ and
  $\int_{0}^{1} \left[\xi f_Y\{F_Y^{-1}(z);\boldsymbol{\eta} \}
    C(z;\boldsymbol{\eta},\boldsymbol{\psi})\right]^{-1}dz = \infty$.
\end{theorem}
A proof is given in Appendix. Theorem~\ref{th:th1} gives the
characterization of the class of distribution functions $F_{V}$ from
which can be constructed a new class of models,
$F_Y^{-1}(V;\boldsymbol{\eta})$, that have tail index $\xi \in
\mbbR$. Under the assumption of $F_Y=F$ and
$\boldsymbol{\eta}=\boldsymbol{\lambda}$, i.e., the case in the
examples of \S~2.1, we obtain the following.

\begin{corollary}
  Let $V$ be a random variable as in Theorem~\ref{th:th1}. Then
  $F^{-1}(V;\boldsymbol{\lambda})$ has a tail index $\xi \in \mbbR$ if
  and only if the distribution function of $V$ is given by
  \begin{equation}
    F_{V}(v;\boldsymbol{\theta}) = 1-\exp\left\{-\int_{0}^{v} \left[\xi (1-z)^{1+\xi} \int
        \frac{s(z;\boldsymbol{\lambda,\boldsymbol{\psi}})}{(1-z)^{1+\xi}}dz\right]^{-1}dz\right\},
    \quad 0 \leq v \leq 1,
    \label{eq:VdfGP}
  \end{equation}
  where $s$ is a real-valued function with $\lim_{z \rightarrow 1}
  s(z;\boldsymbol{\lambda},\boldsymbol{\psi}) = 1$ and $\int_{0}^{1}
  [\xi (1-z)^{1+\xi} \int
    \frac{s(z;\boldsymbol{\lambda,\boldsymbol{\psi}})}{(1-z)^{1+\xi}}dz]^{-1}dz=\infty$.
\end{corollary}
Any real-valued function $s$ with the specific properties of
Theorem~\ref{th:th1} would give rise to a valid distribution function
$F_V$. As an example, consider the real-valued function
$s(z;\boldsymbol{\lambda}, \boldsymbol{\psi} )$ where
$\boldsymbol{\psi}=\kappa \in (0,\infty)$, given by
\[
s(z;\boldsymbol{\lambda},\boldsymbol{\psi})=z^{-\kappa}\left\{z^\kappa
  +\kappa-1 + z^{\kappa+1}\xi - z(\kappa+\xi)\right\} /\left\{\kappa
  \xi (z-1)\right\}, \quad 0 \leq z\leq 1.
\]

\noindent Then, equation~(\ref{eq:VdfGP}) yields
\begin{IEEEeqnarray}{rCl}
  F_V(v;\boldsymbol{\theta})=1 - \exp\left\{-\int_{0}^{v}\left[\xi
      (1-z)^{1+\xi}\left\{\frac{(1-z^{\kappa})}{\kappa \xi
          z^{\kappa-1} (1-z)^{1+\xi}} +
        c\right\}\right]^{-1}dz\right\},\quad c\in \mbbR.
\label{eq:VdfGPex3}
\end{IEEEeqnarray}
When $c=0$, the distribution function $F_{V}(v;\boldsymbol{\theta}) =
v^{\kappa} $ of Example 3 in \S~\ref{sec:pit} is obtained. When $c>0$
and $\xi\geq 0$, equation~(\ref{eq:VdfGPex3}) yields a valid
distribution function $F_V(v;\boldsymbol{\theta})$. However, $c$ has
to be necessarily equal to 0 for $F_V(v;\boldsymbol{\theta})$ to be a
valid distribution function when $\xi<0$.

\subsection{Penultimate approximations}
We have so far presented three examples from a general class of models
that extends the GP distribution by incorporating additional
parameters while preserving the tail index $\xi \in \mbbR$. To
characterise the deviation of the tail behaviour of the extended
models from the GP distribution we examine the penultimate
approximation of the tail index proposed by \cite{smit87}, i.e., we
examine the rate of convergence of the three extended models given in
\S~\ref{sec:pit} to the GP survival function in limit
expression~(\ref{eq:gpdlimit}). Let $W$ be a random variable with
twice differentiable distribution function $G(x)$ and density function
$g(x)$. Denote also the reciprocal hazard function of $W$ by $h_W(x) =
[1-G(x)]/g(x)$. \cite{smit87} shows that for each $u$ and $x>0$ there
exists $y \in [u,u+x h(u)]$ such that
\begin{equation}
\frac{1-G(u+x h(u))}{1-G(u)} = \left\{1 +
  h'(y)x\right\}_{+}^{-1/h'(y)}.
\label{eq:mvt}
\end{equation}
By virtue of expression~(\ref{eq:gpdlimit}) the scaled excess random
variable $\{(W-u)/h(u)\}|W>u$ converges in distribution to the GP
distribution if $\lim_{u\rightarrow x^{G}}h'(u) = \xi \in \mbbR$. This
is one form of the von Mises condition which is a necessary and
sufficient condition for the convergence of the scaled excess of any
random variable, with twice differentiable distribution function, to
the GP distribution. Defining $u_n=G^{-1}(1-1/n)$, the penultimate
approximation to the tail index in equation~(\ref{eq:mvt}) is given in
terms of $n$ by $h'(u_n)$, as $n\rightarrow \infty$.  Moreover, the
rate of convergence to the GP distribution is given by
$O\{|h'(u_n)-\xi|\}$.  \newline

\noindent
Define
$A_{\kappa,\xi}=\frac{(\kappa-1)}{1+1/|\xi|}\left\{|\xi|\big/\Beta(\kappa,1/|\xi|)\right\}^{-|\xi|}$
and let $D_{\kappa,\xi}=(\xi-|\xi|) A_{\kappa,\xi}$ and
$E_{\kappa,\xi}=|\xi| A_{\kappa,\xi}$ for $\xi \neq
0$. Table~\ref{tab:penultimate} shows the leading order terms from the
penultimate approximations of the tail index for the EGP models.  For
$\xi\in [-1,1 ]$, the EGP3 distribution admits the fastest rate of
convergence whereas for $\xi \in [-1,1]^c$ the EGP1 distribution has
the fastest rate of convergence among the extended
models. Irrespective of the value of $\xi$ the EGP2 distribution has
the slowest rate of convergence. Explicitly, for $\xi \neq 0$ and
$\xi=0$, the rate of convergence for the EGP1, EGP2 and EGP3
distributions is of order $\left\{n^{-|\xi|I(\xi>0) }n^{-2|\xi|I(\xi
    <0)},(\log n)^{-1}, n^{-1}\right\}$ and $\left\{(\log
  n)^{-2},(\log n)^{-2}, n^{-1}\right\}$, respectively. Here
$I{\left(\xi\in A\right)}$ denotes the indicator function which
takes the value 1 when $\xi \in A$ and 0 otherwise for any set $A\in
\mbbR$.

\begin{table}[!htbp]
  \caption{Leading
    terms of threshold $u_n$ and penultimate approximations $h'(u_n)$ 
    for Examples~1--3 of \S~\ref{sec:pit}.}
  \label{tab:penultimate}
  \centering     
  \begin{tabular}{c c c c c}
    \hline
    &&&&\\
    Model& $u_n $ & $h'(u_n) (\xi \neq 0)$ & $h'(u_n) (\xi \rightarrow 0)$  \\
    &&&&\\
    \hline
    \text{EGP1}&$(\sigma/\xi)\Big[\left\{\frac{n
        |\xi|}{\Beta(\kappa,1/|\xi|)}\right\}^{\xi}-1\Big] $&
    $ \xi + n^{-|\xi|}D_{\kappa,\xi} - n^{-2|\xi|}E_{\kappa,\xi}$ & $-(\log n)^{-2}(\kappa-1)$\\\\
    \text{EGP2}&$(\sigma/\xi)\left\{n^{\xi}\Gamma(\kappa)^{-\xi}-1\right\}$& 
    $\xi + (\log n)^{-1}(\kappa-1) $& 
    $-(\log n)^{-2}(\kappa-1)$\\\\
    \text{EGP3}&$(\sigma/\xi)\left\{(\kappa n)^{\xi} - 1\right\}$&
    $\xi + n^{-1}(\kappa-1)(\xi - 1)/(2 \kappa)$&
    $-n^{-1} (\kappa-1)/\kappa  $\\\\
    \hline    
  \end{tabular}
\end{table}

\subsection{Statistics using extended GP Models}
We propose the use of the EGP models as alternatives to the GP
distribution for the modelling of the excess random variable
$X-u|X>u$. Specifically, given a random sample $\boldsymbol{x} $ we
model the exceedances $\boldsymbol{x}_{>u}=\{x_i:
x_i>u\}=:(x_1,\hdots,x_{n_{u}})$ with the EGP($\kappa,\sigma,\xi$)
family of distributions. Maximum likelihood is used to estimate the
parameters $(\kappa,\sigma,\xi)$, i.e., maximum likelihood estimates
satisfy
\[
(\hat{\kappa},\hat{\sigma},\hat{\xi}) = \underset{(\kappa,\sigma,\xi)
  \in S}{\mbox{argmax}}~\ell(\kappa,\sigma,\xi|\boldsymbol{x}_{>u})
\]

\noindent where $S = (0,\infty)\times(0,\infty)\times(-\infty,\infty)
$ and $\ell(\kappa,\sigma,\xi|\boldsymbol{x}_{>u})$ denotes the
log-likelihood of the parameters given the observed sequence of
excesses of length $n_{u}=\#\{x_i>u\}$, i.e., for $\xi \neq 0$
\begin{IEEEeqnarray*}{rCl}
  \ell^{\mbox{\tiny{EGP1}}}(\kappa,\sigma,\xi|\boldsymbol{x}_{>u}) &=&
  n_{u}\log\left\{\frac{|\xi|/\sigma}{\mbox{Be}(\kappa,|\xi|^{-1})}\right\}
  +
  (\kappa-1)\sum_{i=1}^{n_{u}} \log\left[1 - \left\{1+\xi (x_{i}-u)/\sigma\right\}_{+}^{-|\xi|/\xi}\right]- \nonumber\\
  &&-(1/\xi+1)\sum_{i=1}^{n_{u}} \log\left\{1+\xi (x_{i}-u)/\sigma\right\}_{+},
  \end{IEEEeqnarray*}
  \begin{IEEEeqnarray*}{rCl}
  \ell^{\mbox{\tiny{EGP2}}}(\kappa,\sigma,\xi|\boldsymbol{x}_{>u}) &=&
  n_{u} \log\left\{\frac{\sigma^{-1}}{\Gamma(\kappa)}\right\} +
  (\kappa-1)\sum_{i=1}^{n_{u}}\log\left[\xi^{-1}\log\left\{1+\xi (x_{i}-u)/\sigma\right\}_{+}\right]-  \nonumber\\
  &&-(1/\xi+1)\sum_{i=1}^{n_{u}} \log\left\{1+\xi
    (x_{i}-u)/\sigma\right\}_{+}, 
  \end{IEEEeqnarray*}
  \begin{IEEEeqnarray*}{rCl}
  \ell^{\mbox{\tiny{EGP3}}}(\kappa,\sigma,\xi|\boldsymbol{x}_{>u}) &=&
  n_{u}\log\left(\kappa/\sigma\right) +
  (\kappa-1)\sum_{i=1}^{n_{u}} \log\left[1 - \left\{1+\xi (x_{i}-u)/\sigma\right\}_{+}^{-1/\xi}\right]- \nonumber\\
  &&-(1/\xi+1)\sum_{i=1}^{n_{u}} \log\left\{1+\xi
    (x_{i}-u)/\sigma\right\}_{+}.
\end{IEEEeqnarray*}
Inference for extreme quantiles is made via the $T$-observation return
level $x_T$ which is defined by the level that is exceeded on average
once every $T$ observations. The $T$-observation return level is the
solution of $\Pr(X>x_T)=1/T$. Under the assumption that the
exceedances above a threshold $u$ are well modelled by the EGP family
of distributions and such that $x_T>u$, the $T$-observation return
level for $\xi \neq 0$ is given by
\begin{IEEEeqnarray*}{rCl} 
  x_{T}^{\text{\tiny{EGP1}}}& =& u+\frac{\sigma}{\xi}\left[\left\{1 -
      \beta^{-1} \left(1 -
        (T\zeta_{u})^{-1},\kappa,|\xi|^{-1}\right)\right\}
    ^{-\xi/|\xi|}-1\right], \\\nonumber\\
  x_{T}^{\text{\tiny{EGP2}}} &=& u+\frac{\sigma}{\xi}\left[\exp
    \left\{\xi \gamma^{-1}\left(\kappa,1-(T
        \zeta_{u})^{-1}\right)\right\} - 1
  \right], \\\nonumber\\
  x_{T}^{\text{\tiny{EGP3}}} &=& u+\frac{\sigma}{\xi} \left[\left\{ 1
      - (1 - (T\zeta_u)^{-1/\kappa}) \right\}^{-\xi} - 1 \right],
\end{IEEEeqnarray*}                                                 
where $\zeta_{u}=\Pr(X>u)$. Return level estimates are obtained by
substituting the parameter values by their maximum likelihood
estimates whereas standard errors and confidence intervals are derived
by the delta method or from the profile likelihoods of the
parameters. \newline

\noindent Aside from the model fitting of the exceedances with the EGP
family of distributions, additional diagnostics for the GP
distribution can be obtained. In particular, extra insight about the
convergence in expression~(\ref{eq:gpdlimit}) can be sought from the
EGP models by testing the statistical hypothesis
\begin{IEEEeqnarray}{C} 
  H_{0}:\left(\kappa,\sigma,\xi\right) \in S_{0}
  \quad
  \mbox{vs} \quad H_{1}:\left(\kappa,\sigma,\xi\right) \in S_{1},
  \label{eq:stattest}
\end{IEEEeqnarray}
where $S_{0}= \left\{(1,\sigma,\xi):\sigma\in \mbbR_{+},\xi\in
  \mbbR\right\}$ and $S_{1}=
\left\{(\kappa,\sigma,\xi):\kappa\in\mbbR_{+}\setminus \{1\},\sigma\in
  \mbbR_{+},\xi\in \mbbR\right\}$.\ Given the sample of excesses
$\boldsymbol{x}_{>u}$, the generalised log-likelihood ratio test statistic
reads
\begin{IEEEeqnarray}{rCl}
  \Lambda_{n_{u}}(\boldsymbol{x}_{>u}) &=& 2\left[\sup\left\{
      \ell(\kappa,\sigma,\xi|\boldsymbol{x}_{>u}):(\kappa,\sigma,\xi) \in
      S\right\} - \sup\left\{
      \ell(\kappa,\sigma,\xi|\boldsymbol{x}_{>u}):(\kappa,\sigma,\xi) \in
      S_{0}\right\}\right],
\end{IEEEeqnarray}
where $S=S_{0}\cup S_{1}$. From asymptotic likelihood theory as
$n_{u}\rightarrow \infty$, $\Lambda_{n_{u}}$ converges in distribution
to the chi-squared with 1 degree of freedom under $H_{0}$. Therefore,
tests of the statistical hypothesis~(\ref{eq:stattest}) can be made on
the basis of the asymptotic distribution of
$\Lambda_{n_{u}}$. Moreover, the limit expression~(\ref{eq:gpdlimit})
suggests that if the GP distribution is a reasonable model for the
observed exceedances above a threshold $u'$, then exceedances above a
higher threshold $u''\geq u'$ should also follow the GP
distribution. This argument suggests plotting $\hat{\kappa}$ against
$u$ and selecting the threshold as the lowest possible value at which
$\hat{\kappa}$ is not significantly different from 1 and the estimated
modified scale and tail index are constant for all $u''>u'$.

\section{Simulation Study}
\label{sec:simulation}
We illustrate the impact of the extended models on the tail estimation
using normal simulated data. All comparisons are based on the root
mean square error (RMSE) performance of a range of estimated extreme
quantiles using various sample sizes for the
simulations. Specifically, for each distribution 10000 samples of size
$n=100,1000,10000$ were generated. The GP, EGP1 and EGP2 distributions
were fitted to the exceedances of each sample above a range of $N$
equally spaced thresholds $u_{1},...,u_{N}$, with
$u_{1}=\Phi^{-1}(1/n)$ and $u_{N}=\Phi^{-1}(1 - 30/n)$. Results
obtained from the EGP3 model are not shown as they are similar to the
EGP1 and EGP2 models. This grid was chosen such that $u_1$ and $u_N$
correspond approximately to the minimum possible threshold, i.e., all
data are above $u_1$, and $u_N$ is the threshold above which 30 data
points are observed on average, respectively. At each threshold we
computed Monte Carlo estimates of the RMSE of the $T$-observation
return level estimate. For each sample size $n$ used in the simulation
study, we chose two different values of $T$, given by $T_{i}/n=1.5,5$,
for $i=1,2$, corresponding to short and long extrapolations. \newline

\begin{figure}[htpb!]
  \centering
  \includegraphics[scale=1,trim=20 50 20 50]{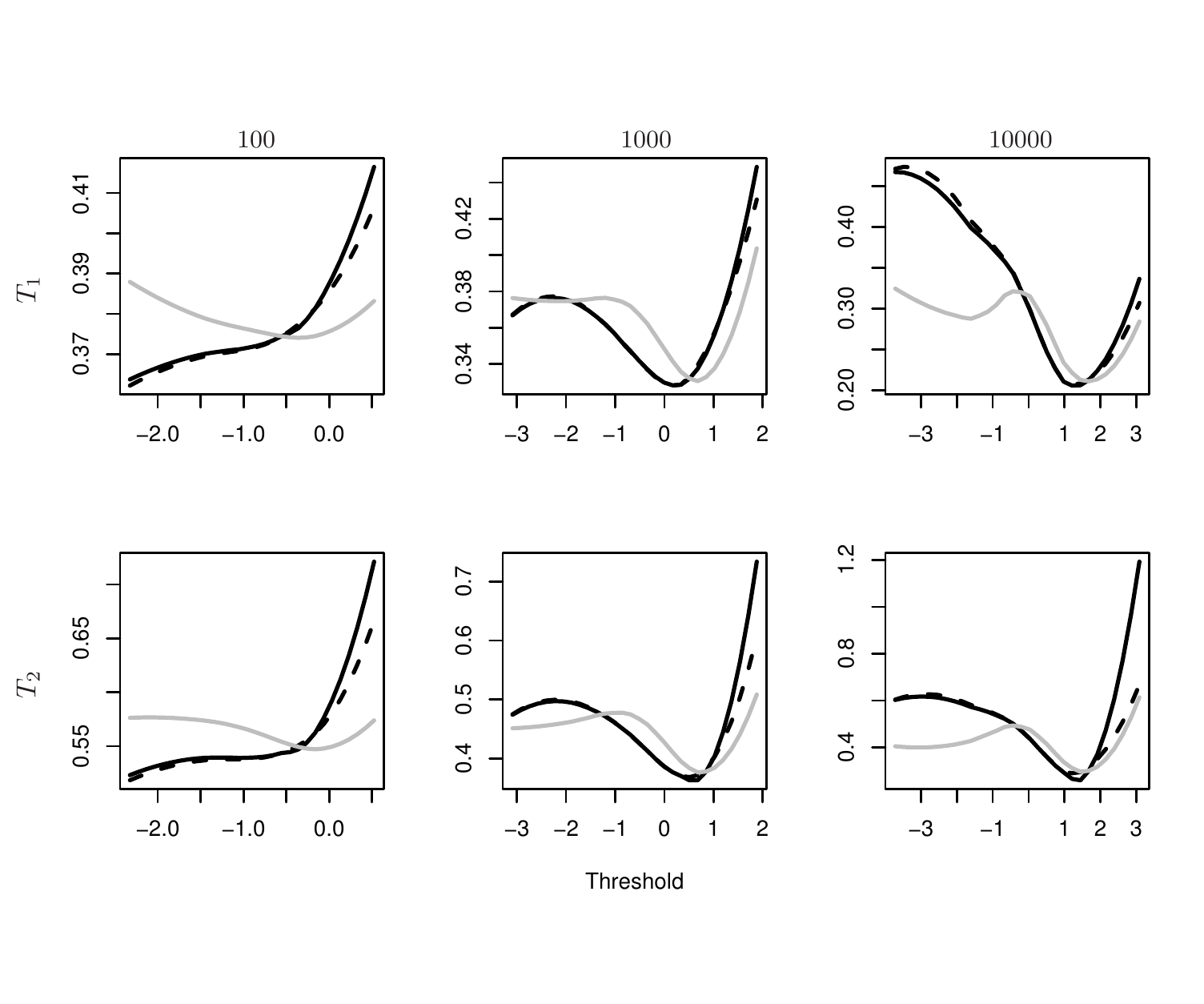}
  \caption{RMSE for the $T_1$-(top row) and $T_2$-observation (bottom
    row) return level estimates obtained from EGP1 (solid black), EGP2
    (dashed black) and GP (grey). Columns correspond to the three
    different sample sizes $n=100,1000,10000$ (from left to right).}
  \label{fig:rmse_normal}
\end{figure}

\noindent Figure \ref{fig:rmse_normal} shows the RMSE output of the
simulation study for the normal simulated data. Results illustrate
improvement in inference using the EGP models over the GP model for
both return level estimates and each sample size as the minimum RMSE
is attained for the two EGP models, with their performance being
almost indistinguishable at this value. More precisely, this
improvement is largest in the small sample case ($n=100$) where the
optimal choice of the threshold according to the lowest RMSE is
$u_1=\Phi^{-1}(1/\{100\})$. This illustrates the advantage of fitting
the EGP models to the whole data in small sample size cases instead of
the GP distribution. For the $T_1$-observation return level in the
$n=100$ case, the EGP estimates yield higher bias and lower variance
than the GP estimates whereas for any other combination of sample size
and return level, the EGP estimates have lower bias and either
slightly lower or higher variance at the threshold where the minimum
RMSE occurs. From Table~\ref{tab:optimal_u} we also have that as the
sample size increases, the absolute difference of the corresponding
optimal thresholds and RMSE of the EGP distributions and the GP
distribution diminishes. This is an expected phenomenon which is
justified by the validity of the asymptotics of extreme value theory
as sample size increases.
\begin{table}[htbp!]
  \caption{Optimal thresholds for the estimation of the 
    return levels for each sample size. }  
  \centering
  \begin{tabular}{c|ccc|ccc}
    \hline\hline
    &\multicolumn{3}{c|}{$T_1$}&\multicolumn{3}{c}{$T_2$}\\
    \hline\hline
    $u \big \backslash n$&100 & 1000 &10000 &
    100  & 1000 &10000\\
    \hline
    $u_{\text{EGP}}$&-2.32&0.05&1.48& -2.32 &
    0.51 & 1.44\\    
    $u_{\text{GP}}$&-0.33&0.45&1.70&-0.23 & 
    0.68&1.44\\
    \hline\hline
  \end{tabular}
  \label{tab:optimal_u}
\end{table}

\begin{figure}[!htpb]
  \centering
  \includegraphics[trim=20 40 20 20]{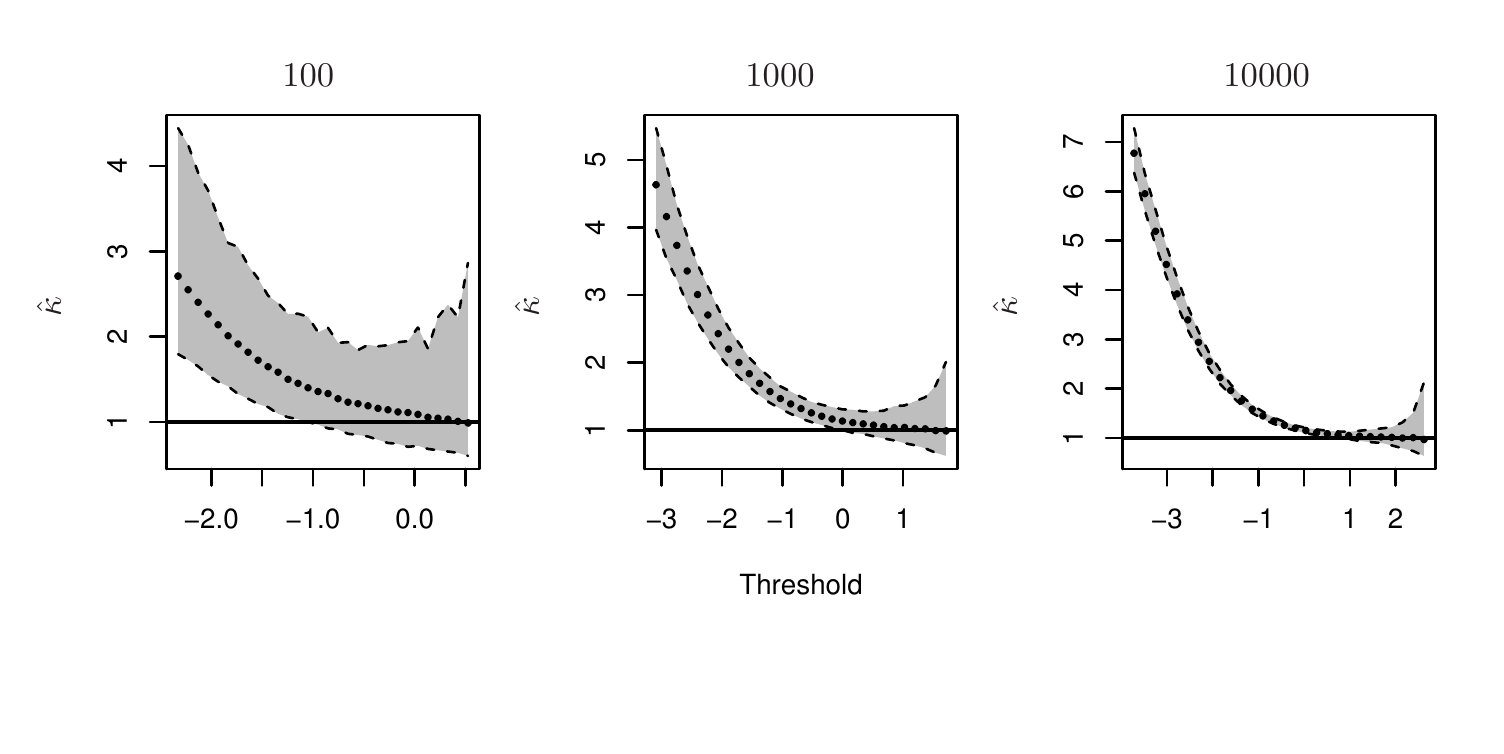}
  \caption{Monte Carlo estimates (median) of parameter $\kappa$ (black
    dots) plotted against the threshold for each sample
    size. Grey-shaded areas correspond to the 95\% pointwise equal
    tail confidence intervals.}
  \label{fig:kappahat}
\end{figure}

\noindent
Figure \ref{fig:kappahat} shows the Monte Carlo estimates as well as
the estimated uncertainty of the shape parameter $\kappa$ from the
EGP2 model plotted against the threshold for all sample sizes. Note
also that the estimates obtained from the EGP1 model are close to the
EGP2 estimates and therefore are not shown here. All graphs illustrate
the same feature, i.e., $\hat{\kappa}$ stabilises around the value 1
as the threshold $u$ increases. Additionally, the minimum thresholds
at which the value 1 is inside the sampling distribution of $\kappa$
are similar to the optimal thresholds of Table~\ref{tab:optimal_u} for
the GP model, denoted by $u_{\text{GP}}$. This feature demonstrates
the usefulness of this plot as an additional diagnostic for the GP
modelling framework. The 95\% pointwise confidence intervals are
largest for small and large threshold values. This feature is
explained by the greater dependence of parameters $\xi$ and $\kappa$
at low threshold values (revealed by the profile likelihood plots of
$\xi$ and $\kappa$ that are not shown here) and the few data points at
high threshold values.

\clearpage
\section{Applications}
\label{sec:apps}
\subsection{River Nidd Data}
\label{sec:nidd}
\begin{figure}[htpb!]
  \centering
  \includegraphics[trim= 20 20 20 10]{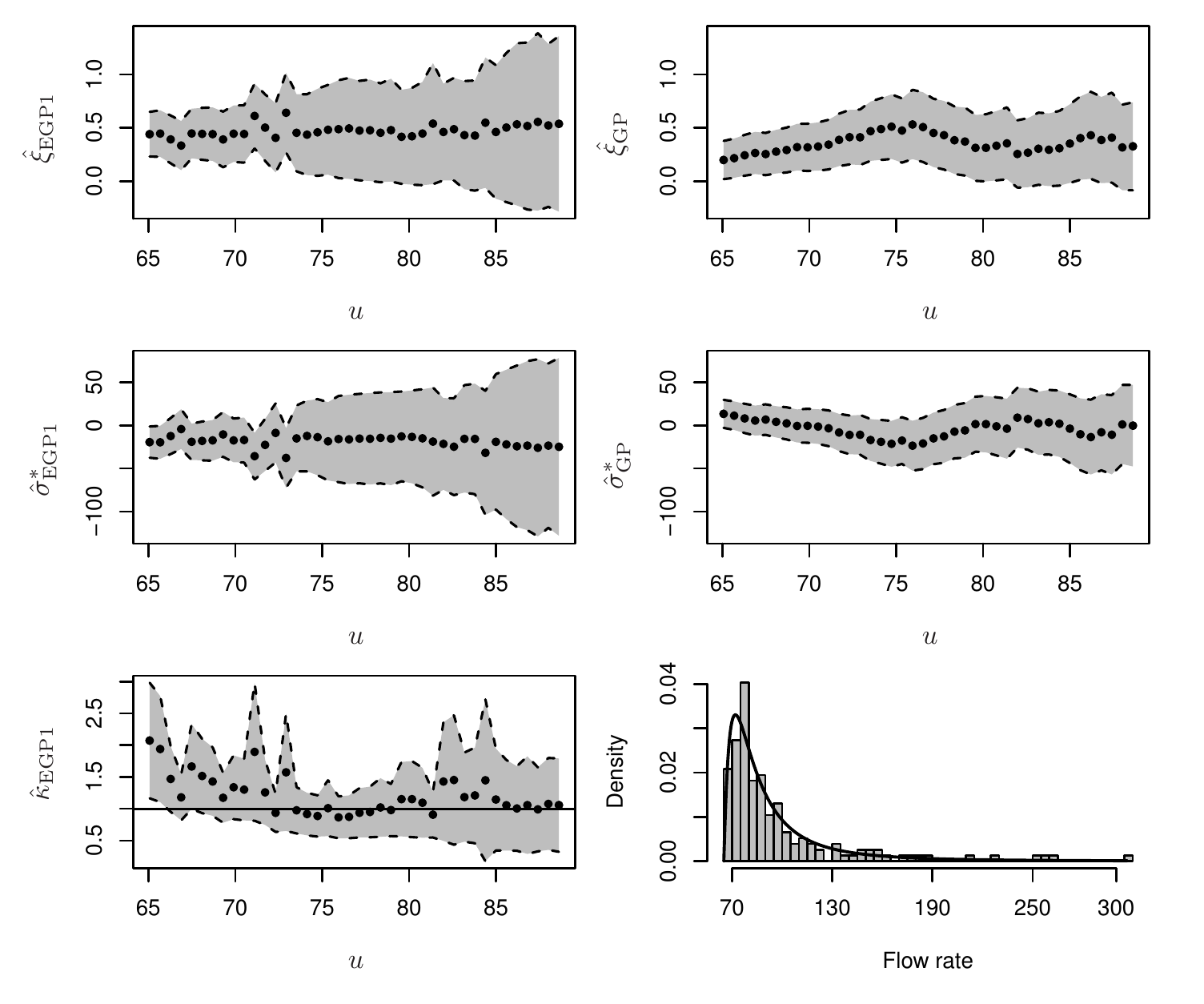}
  \caption{Maximum likelihood estimates and 95\% pointwise equal-tail
    confidence intervals of tail index, modified scale and shape
    parameter $(\xi,\sigma^{*},\kappa)$ based on asymptotic normality:
    EGP1 (left column), GP (right column). Bottom right graph shows
    the histogram of the River Nidd data along with the estimated
    density (black solid line) from the EGP1 model fitted to the
    exceedances above $u=65$m$^3$s$^{-1}$.}
  \label{fig:niddres}
\end{figure}

We now analyse 154 exceedances of the threshold 65m$^3$s$^{-1}$ by the
River Nidd at Hunsingore Weir from 1934 to 1969 taken from
\cite{FSR}. This data set constitutes the best known example with
apparent difficulties in threshold selection and the modelling of the
tail using the GP distribution, studied previously by
\cite{hoskwallis87}, \cite{davismit90}, \cite{tancredi06} and
\cite{wadtawn11}. Figure \ref{fig:niddres} shows the parameter
stability plots from the EGP1 (left) and GP (right) models over a grid
of thresholds $65.08,\hdots,88.61$ along with the histogram of the
data. Threshold selection from the GP model based on the stability of
the tail index and modified scale parameters is not
straightforward. In contrast, the tail index and modified scale
estimates from the EGP1 model appear to be stable over the plotted
range of thresholds. Hence we select $u=65$m$^3$s$^{-1}$ (all data
points) for the fit of the EGP1 distribution.
Moreover, the fact that
$(\hat{\kappa}_{\text{EGP1}},\hat{\sigma}^{*}_{\text{EGP1}},\hat{\xi}_{\text{EGP1}})$
stabilise to values around $(1,-16,0.46)$ for the threshold values
above 74 suggests that any threshold in this region is reasonable for
the GP distribution. However, small deviations of
$\hat{\kappa}_{\text{EGP1}}$ from the value 1 in this threshold region
seem to have an impact on the stability of the GP estimates and the
lowest threshold where $\hat{\kappa}$ is very close to 1 is
75.3m$^3$s$^{-1}$. This finding is also consistent with that of the
\cite{wadtawn11} approach where they choose the value of
75m$^{3}$s$^{-1}$. We thus select and $u=75.3$m$^3$s$^{-1}$ for the GP
distribution. Note also that the tail index and modified scale
estimates from the EGP1 fitted above $u=65$m$^3$s$^{-1}$ $(0.44,-19)$
are similar to those obtained from the GP fitted above
$u=75.3$m$^3$s$^{-1}$ $(0.48,-17)$.\newline

\begin{figure}[htbp!]
  \centering
  \includegraphics[scale=0.6,trim=10 10 10 10]{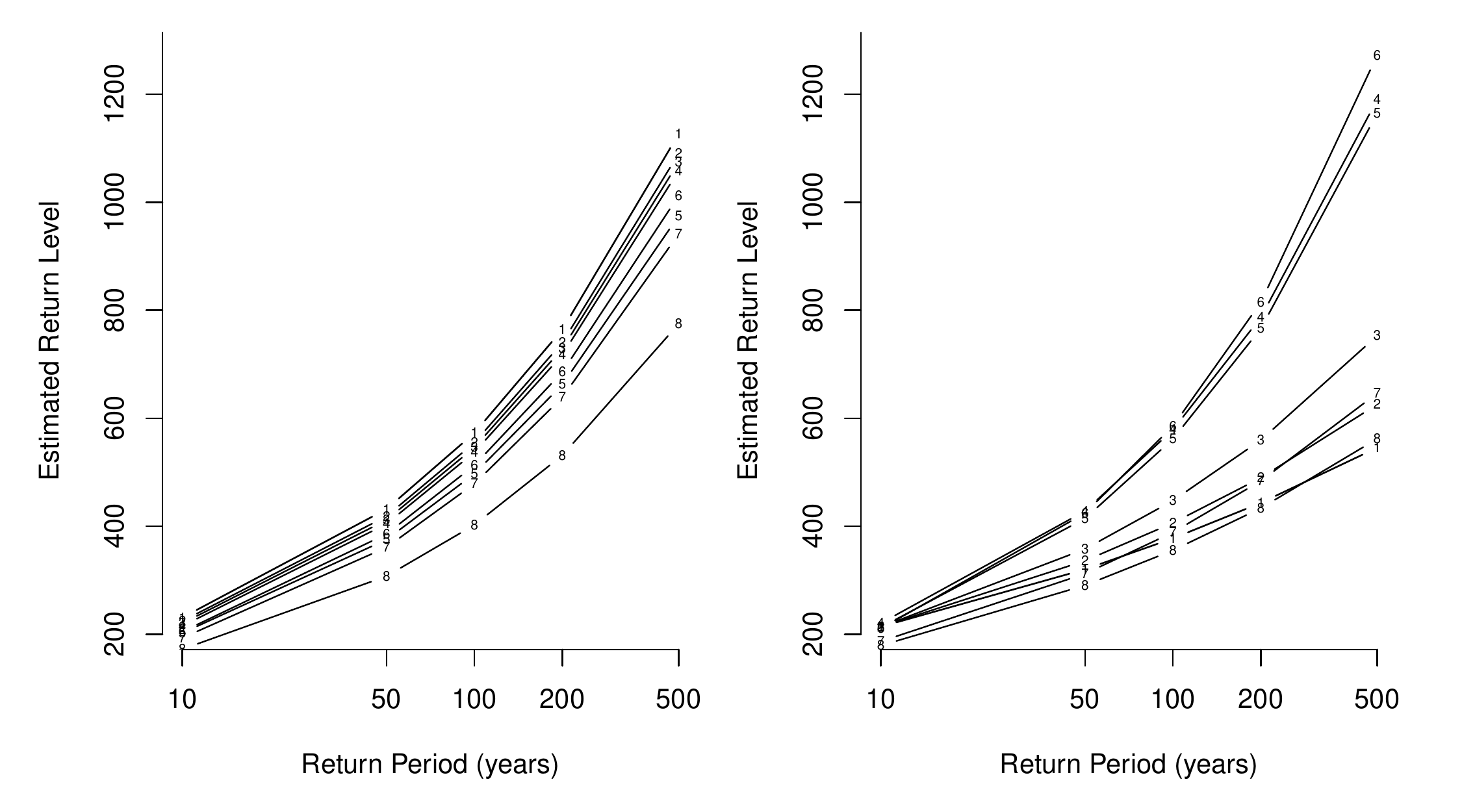}
  \caption{Estimates of 10-, 50-, 100-, 200- and 500-observation
    return level obtained from fitted EGP1 (left) and GP (right)
    models above the range of thresholds $65.08,\hdots,82.6$ coded
    here by numbers $1,\hdots,8$ respectively. Numbers 1 and 5
    correspond to thresholds $65.08$ and $75.3$ respectively.}
  \label{fig:nidd_ret_levs}
\end{figure}

\noindent
To assess the impact on extrapolation, we look at the stability of
return level estimates with respect to the choice of the
threshold. Figure~\ref{fig:nidd_ret_levs} shows return level estimates
obtained from the EGP1 and GP models on the same grid of
thresholds. Clearly, inference made on the basis of the EGP1 model
yields much more stable results in comparison with the GP
model. Return level estimates obtained from the EGP1 model gradually
decrease with increasing threshold whereas estimates obtained from the
GP model vary irregularly. This feature illustrates that the choice of
threshold is less important for the Nidd data while using the EGP
class of distributions.

\subsection{Pharmaceutical Application}
\label{sec:pharma}
We now return to the analysis of the residual bilirubin data shown in
Figure~\ref{fig:pharma_data}.\ As already mentioned in
\S~\ref{sec:intro}, the identification of liver toxic drugs is a
multivariate extreme value problem in which the joint occurrence of
extremes of residual bilirubin and other laboratory variables must be
well modelled. However, as any multivariate extreme value analysis
necessitates, the marginal extremes of these variables have to be
modelled first.\ \cite{texmex} analysed the extremes of all laboratory
variables taken from the same dataset with the GP modelling approach
of \cite{davismit90}, taking the threshold as the 70\% quantile of the
data. They found dose response relationships for all liver related
laboratory variables other than residual bilirubin, justified by GP
models with scale or tail index parameters linear in dose. Our primary
objective in this analysis is to use the EGP1 distribution of
\S~\ref{sec:pit} to model the extremes of the residual bilirubin and
to test for relationship with dose. Using the EGP models of
\S~\ref{sec:pit} allows the inclusion of more data points which might
reveal evidence of relationship between residual bilirubin and dose,
missed by \cite{texmex}. To assess the relationship of residual
bilirubin with dose we use generalised likelihood ratio tests between
models that have dose dependent parameters and models with the same
parameters across doses. The practice of pooling parameters and more
specifically of the tail index in the extreme value modelling
framework can be found in various applications including
\cite{coletawn90,cooetal07} and \cite{davpadrib} to name but a
few. \newline

\begin{figure}[!htpb]
  \centering
  \includegraphics[scale=1,trim= 2 2 2 2]{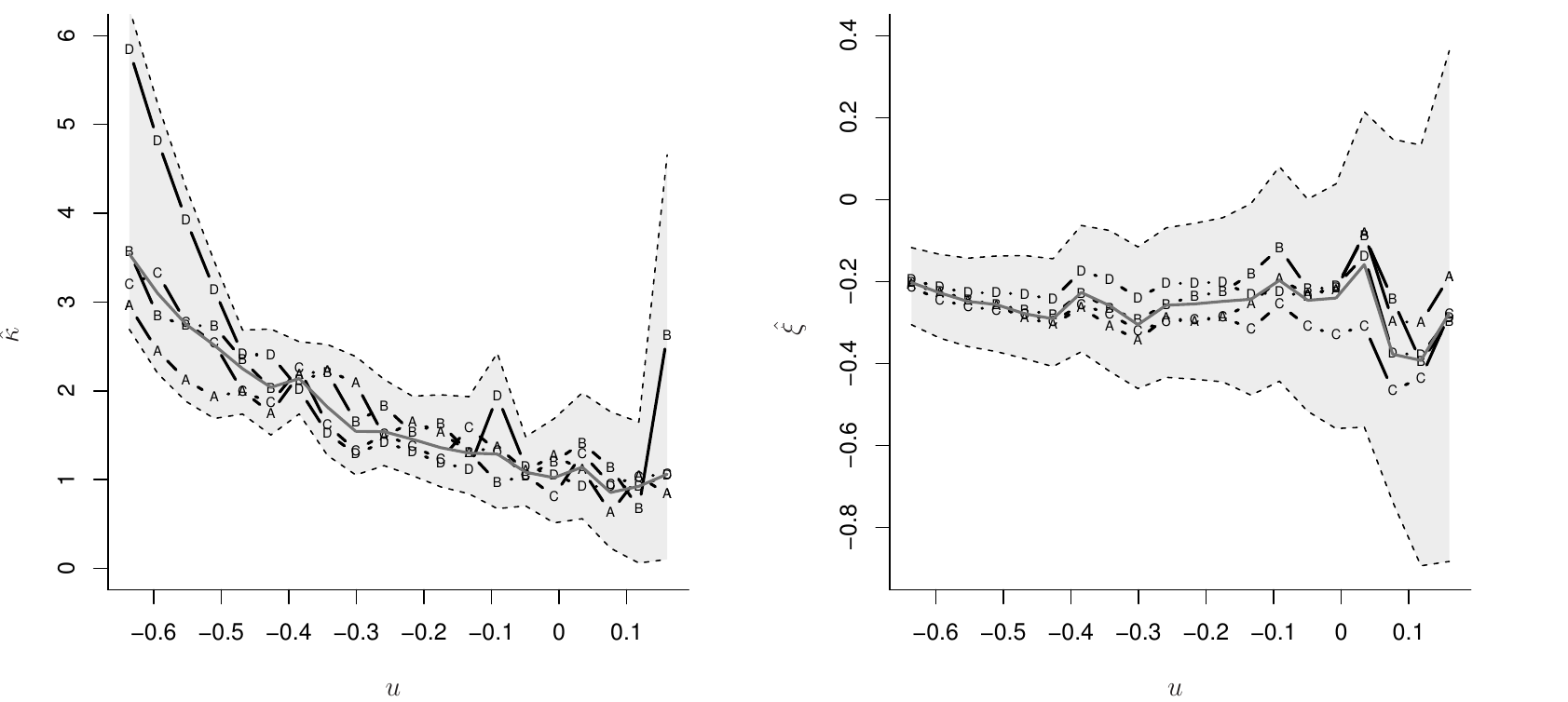} 
  \caption{Left: estimated shape parameter for dose levels $A$, $B$,
    $C$, $D$ (black lines) and under the assumption of common shape
    across doses (dark grey line). Right: estimated tail index for
    dose levels $A$, $B$, $C$, $D$ under common shape across doses
    (black lines) and under the assumption of common shape and common
    tail index across doses (dark grey). Light grey areas correspond
    to a set containing all 95\% pointwise confidence intervals based
    on asymptotic normality for the parameter estimates shown with
    black lines. The set is constructed by the minima (lower boundary)
    and maxima (upper boundary) of the confidence intervals.}
  \label{fig:kappaxihat_pharma}
\end{figure}

\noindent Let $X^{j}-u|X^{j}>u$ be the excesses of the residual
bilirubin variable over the threshold $u$ at dose $j=A, B, C, D$. We
initially fit the EGP1 model to the excesses over thresholds ranging
from -0.65 ($1\%$) to 0.15 ($77\%$) by allowing separate shape, scale
and tail index parameters for each dose, i.e., $X^{j}-u|X^{j}>u \sim$
\text{EGP1}$(\kappa_{j},\sigma_{j},\xi_{j})$, for dose $j$. The
numbers in brackets are the corresponding sample quantiles of the
combined data. The left plot of Figure~\ref{fig:kappaxihat_pharma}
shows the maximum likelihood estimates
$\hat{\kappa}_{A},\hdots,\hat{\kappa}_{D}$ over the threshold
values. A feature revealed from this graph is that the estimated shape
parameters appear to be similar across the doses for thresholds
greater than $-0.51$. This is also supported by the generalised
likelihood ratio test of the hypothesis $H_{0}:
(\kappa_{A},\hdots,\kappa_{D})\in Q $ vs $H_{1}:
(\kappa_{A},\hdots,\kappa_{D})\in Q^{c}$, where
$Q=\{(\kappa_A,\hdots,\kappa_D) \in
\mbbR_{+}^{4}:\kappa_A=\hdots=\kappa_D\}$ and $Q^{c}$ is the
complement of the set $Q$. Specifically, the generalised likelihood
ratio test failed to reject the null hypothesis at all thresholds
other than the threshold values below -0.51. Thus, we proceed to the
analysis of the bilirubin data with the estimated common shape
parameter shown with the dark grey line in the left plot of
Figure~\ref{fig:kappaxihat_pharma}. The right plot of
Figure~\ref{fig:kappaxihat_pharma} shows the maximum likelihood
estimates $\hat{\xi}_{A}, \hdots, \hat{\xi}_{D}$ under the assumption
of common shape across doses. In this case, the generalised likelihood
ratio test failed to reject the null hypothesis of common tail index
over dose at all thresholds. We found that the simplest model selected
by generalised likelihood ratio tests is with common shape, scale and
tail index parameters for all doses. We also found similar results
regardless of the order according to which the pooling of parameters
was conducted. This suggests that there is no evidence of relationship
between the residual bilirubin and dose for all thresholds greater
than -0.51, at the significance level of 5\%. However, for thresholds
below $-0.51$ there is evidence of a relationship with dose as
indicated by the significant increase in the shape parameter estimate
for dose $D$. This change indicates larger quantiles for dose $D$ than
for the other doses. \newline

\begin{figure}[htpb!]
  \centering \includegraphics{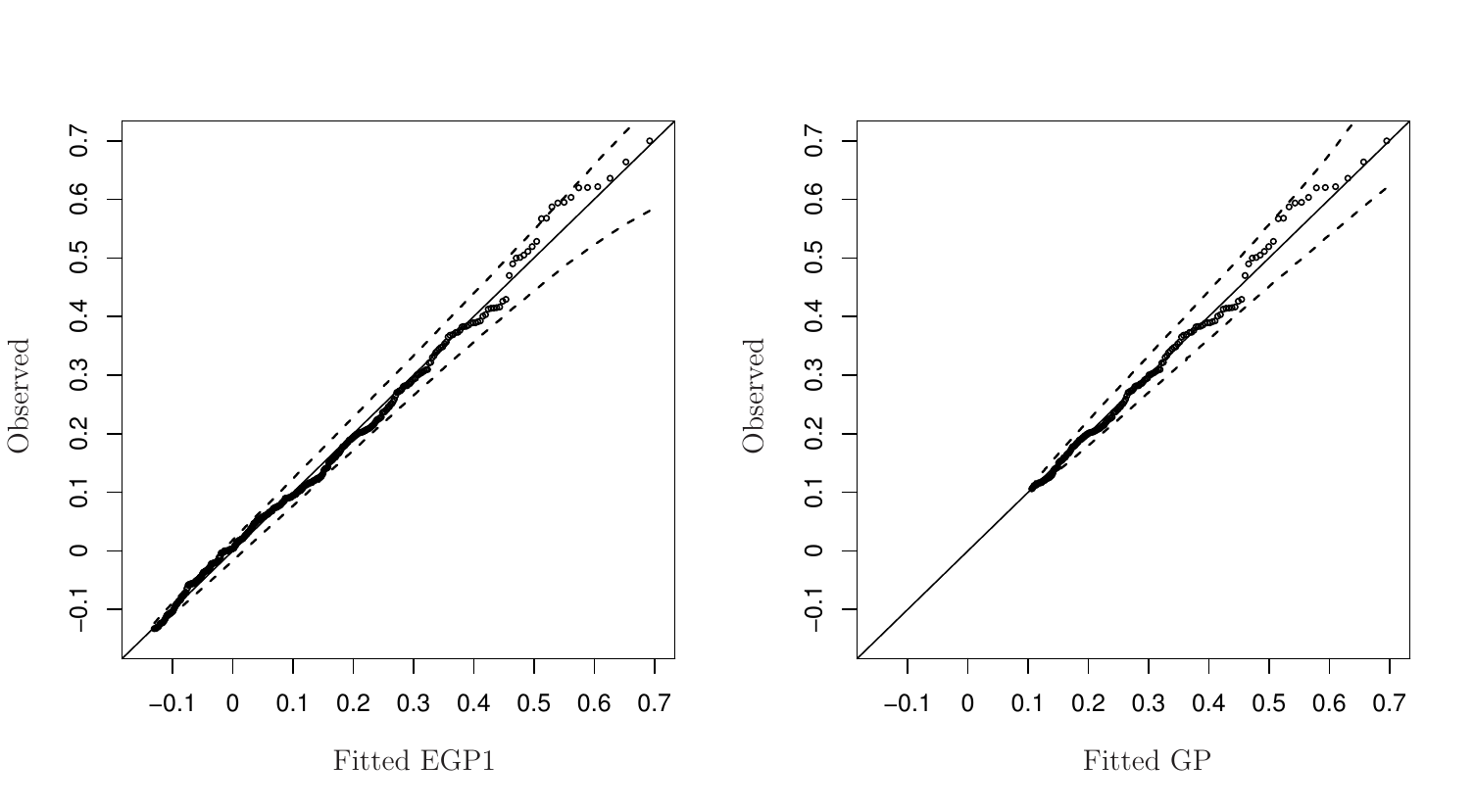}
  \caption{Quantile-quantile plots to assess the fit to the
    exceedances of the EGP1 (left) and GP (right) models. Dashed lines
    show the 95\% pointwise tolerance intervals.}
  \label{fig:QQ}
\end{figure}

\noindent Figure~\ref{fig:QQ} shows the quantile-quantile plots for
the EGP1 and GP models with common shape, scale and tail index
parameters among doses, fitted to the threshold exceedances above 0.10
($30\%$) and -0.13 ($70\%$), respectively. The parameter estimates
obtained from the EGP1 and GP fits are
$(\hat{\kappa}_{\text{EGP1}},\hat{\sigma}_{\text{EGP1}},\hat{\xi}_{\text{EGP1}})
= (1.29,0.25,-0.24)$ and
$(\hat{\sigma}_{\text{GP}},\hat{\xi}_{\text{GP}}) = (0.21,-0.27)$
respectively. Their corresponding standard errors are
$(0.09,0.02,0.05)$ and $(0.01,0.04)$. For the GP model we used
\cite{texmex} choice of the 70\% quantile which is consistent with the
stability of the parameter estimates. For both models, the fit is good
as the majority of the observed data points lie within the 95\%
pointwise tolerance intervals.
\newline
\noindent The best fitting EGP1 model has $\hat{\kappa}$ significantly
different from $1$, and hence provides evidence of a departure from
the GP distribution at the selected threshold. However, above the
respective thresholds used to fit the two models there is no apparent
difference in the quality of the fits. The finding of no evidence of a
dose effect in the EGP models is identical to findings of the previous
GP analysis. Despite this failure to identify a dose effect for
thresholds above $-0.51$, we believe our analysis offers considerable
benefits. Specifically, due to being able to substantially lower the
threshold used relative to the GP analysis, larger sample sizes are
used and thus the power of a test for dose effects in the residual
bilirubin data is increased.

\subsection*{Acknowledgments}
I. Papastathopoulos's work was carried under funding from
Astrazeneca.\ We would particularly like to thank Harry Southworth of
Astrazeneca for helpful discussions, suggestions and constructive
comments on the analysis of the bilirubin data of \S~\ref{sec:pharma}
and Ivar Struijker Boudier for carrying out and validating some of the
numerical calculations of \S~\ref{sec:simulation}.

\appendix
\section{Proof of Theorem 1}

Assume that $F_{V}$ can be represented by equation~(\ref{eq:Vdf}). Let
$K(x;\boldsymbol{\eta},\boldsymbol{\theta})$ and
$k(x;\boldsymbol{\eta},\boldsymbol{\theta})$ be the distribution
function and density function of the transformed variable
$W=F_Y^{-1}(V;\boldsymbol{\eta})$. Differentiability of $f_Y$ and
$f_V$ implies that $W$ will have tail index $\xi \in \mbbR$ if the
derivative of the reciprocal hazard function of $W$,
$h_W'(x;\boldsymbol{\eta},\boldsymbol{\theta})=
d/dx\left[\{1-K(x;\boldsymbol{\eta},\boldsymbol{\theta})\}/
  k(x;\boldsymbol{\eta},\boldsymbol{\theta})\right]$, equals $\xi$ as
$x \rightarrow x^{K}
=\sup\left\{x:K(x;\boldsymbol{\eta},\boldsymbol{\theta})<1\right\}$
(Von Mises' condition). We have
\begin{IEEEeqnarray*}{rCl}
  h_W'(x;\boldsymbol{\eta},\boldsymbol{\theta})&=&
  \xi C'\left\{F_Y(x);\boldsymbol{\eta},\boldsymbol{\psi}\right\}\\\\
  &=& \xi s\left\{F_Y(x);\boldsymbol{\eta},\boldsymbol{\psi}\right\}\\\\
  &\rightarrow& \xi \in \mbbR,\quad \text{as $x\rightarrow x^{K}$}.
\end{IEEEeqnarray*}
To prove the converse, we assume that the random variable $W$ has tail
index $\xi$, i.e.,\newline $\lim_{x\rightarrow
  x^{K}}h_W'(x;\boldsymbol{\eta},\boldsymbol{\theta})=\xi$. In other
words, there exists a real-valued function $s: \mbbR \rightarrow
\mbbR$ with $\lim_{x\rightarrow x^{F_Y}}s\left\{
  F_Y(x);\boldsymbol{\eta},\boldsymbol{\theta} \right\}= 1$ such that
$h_W'(x;\boldsymbol{\eta},\boldsymbol{\theta})=~\xi
s\{F_Y(x);\boldsymbol{\eta},\boldsymbol{\theta}\}$. Writing
$h_W(x;\boldsymbol{\eta},\boldsymbol{\theta})=
h_{V}\{F_Y(x;\boldsymbol{\eta});\boldsymbol{\theta}\}/f_Y(x;\boldsymbol{\eta})$
we have
\[
h_{V}'\{F_Y(x;\boldsymbol{\eta});\boldsymbol{\theta}\} -
\frac{f_Y'(x;\boldsymbol{\eta})}{f_Y^{2}(x;\boldsymbol{\eta})}
h_{V}\{F_Y(x;\boldsymbol{\eta});\boldsymbol{\theta}\} = \xi
s\{F_Y(x);\boldsymbol{\eta},\boldsymbol{\theta}\}.
\]
The solution of this first order linear differential equation is given
by
\[
h_{V}\{F_Y(x;\boldsymbol{\eta});\boldsymbol{\theta}\} = \xi
f_Y(x;\boldsymbol{\eta}) \int
s\{F_Y(x);\boldsymbol{\eta},\boldsymbol{\theta}\} dx,
\]
which is a separable differential equation with solution
\[
F_V\{F_Y(x;\boldsymbol{\eta});\boldsymbol{\theta}\} =
1-\exp\left\{-\int_{0}^{F_Y(x;\boldsymbol{\eta})}\frac{dF_Y(t;\boldsymbol{\eta})}{\xi
    f_Y(t;\boldsymbol{\eta}) \int
    s\{F_Y(t);\boldsymbol{\eta},\boldsymbol{\theta}\} dt}\right\}.
\]
Under the change of variable $z=F_Y(t;\boldsymbol{\eta})$, we have
\begin{equation}
  F_V(v;\boldsymbol{\theta}) =
  1-\exp\left\{-\int_{0}^{v}\frac{dz}{\xi
      f_Y\{F_Y^{-1}(z);\boldsymbol{\eta}\} \int
      \frac{s ( z;\boldsymbol{\eta},\boldsymbol{\theta})}{f_Y\{F_Y^{-1}(z);\boldsymbol{\eta}\}} dz}\right\},
\label{eq:proof1}
\end{equation}
where $v=F_Y(x;\boldsymbol{\eta})$. By assumption
$\boldsymbol{\theta}$ is an at most $d$-dimensional vector of
parameters. Hence equation~(\ref{eq:proof1}) implies the existence of
a $(d-m)$-dimensional vector of parameters
$\boldsymbol{\psi}\in\Psi\subseteq \mbbR^{d-m}$ such that
expression~(\ref{eq:proof1}) can be written as
\begin{equation*}
  F_V(v;\boldsymbol{\theta}) =
  1-\exp\left\{-\int_{0}^{v}\frac{dz}{\xi
      f_Y\{F_Y^{-1}(z);\boldsymbol{\eta}\} \int
      \frac{s ( z;\boldsymbol{\eta},\boldsymbol{\psi})}{f_Y\{F_Y^{-1}(z);\boldsymbol{\eta}\}} dz}\right\},
\end{equation*}
\noindent
and $(\boldsymbol{\psi},\boldsymbol{\eta})$ span $\Theta$.


\end{document}